\documentclass[aps,twocolumn, secnumarabic,nobalancelastpage,amsmath,amssymb,nofootinbib]{revtex4-2}

\usepackage{graphicx}
\usepackage{bm}
\usepackage{bbm}
\usepackage{bbold}
\usepackage{amsmath}
\usepackage{braket}
\usepackage[normalem]{ulem}
\usepackage{soul}
\usepackage{txfonts}
\usepackage{tabularx}
\usepackage{tabularray}
\usepackage[usenames,dvipsnames]{color}
\setstcolor{red}

\newlength\myboxwidth
\begin{document}

\title{Propagation velocity measurements of substrate phonon bursts using MKIDs for superconducting circuits}

\author{Guy Moshel}
\email{gmoshel@campus.technion.ac.il}

\author{Omer Rabinowitz}
\author{Eliya Blumenthal}
\author{Shay Hacohen-Gourgy} 
\affiliation{Department of Physics, Technion - Israel Institute of Technology, Haifa 32000, Israel}
\date{\today }

\begin{abstract}
High-energy bursts in superconducting quantum circuits from various radiation sources have recently become a practical concern due to induced errors and their propagation in the chip. The speed and distance of these disturbances have practical implications. We used a linear array of multiplexed MKIDs on a single silicon chip to measure the propagation velocity of a localized high-energy burst, introduced by driving a Normal metal-Insulator-Superconductor (NIS) junction. We observed a reduction in the apparent propagation velocity with NIS power, which is due to the combined effect of reduced phonon flux with distance and the existence of a minimum detectable QP density in the MKIDs. A simple theoretical model is fitted to extract the longitudinal phonon velocity in the substrate and the conversion efficiency of phonons to QPs in the superconductor.
\end{abstract}

\maketitle
Numerous authors have discussed the role of cosmic radiation and background radioactivity in producing high-energy bursts in superconducting quantum circuits~\cite{mcewen2022resolving,harrington2024synchronous,grunhaupt2018loss}. These bursts are generated by an impact of a high energy particle with the circuit chip. They then spread through the chip and cause decoherence and decay in superconducting components ~\cite{martinis2021saving}. Many approaches has been suggested to mitigate these bursts~\cite{bertoldo2023cosmic,orrell2021sensor,hosseinkhani2017optimal,henriques2019phonon,iaia2022phonon}, and also some experimental works done to characterize and mitigate them~\cite{Swenson2010highSpeed,gordon2022environmental,yelton2024modeling,thorbeck2023two,cardani2021reducing,bargerbos2023mitigation}. In this work we supplement these efforts by accurately measuring the propagation velocity of the bursts and consequently the effective time it takes a burst at a certain distance to cause an error.

The study of burst dynamics is made difficult by their rarity and stochastic nature, with an average occurrence rate of approximately one event per ten seconds per square centimeter. This makes it impossible to perform repeatable and deterministic experiments, which are necessary for systematic investigations. To overcome this, we generate artificial high-energy bursts by controllably injecting phonons into the substrate using a normal metal-insulator-superconductor (NIS) junction~\cite{ullom2002physics,blonder1982transition}. This method was utilized previously~\cite{patel2017phonon}. When a sufficiently high drive signal is applied to the junction, quasi-particles (QPs) from the normal metal are transferred to the superconductor. These QPs then recombine and generate pair-breaking phonons in the substrate which propagate as though emanating from a cosmic ray impact. This allowed us to perform high-accuracy and high temporal resolution measurements.
Although the energies of phonons generated by cosmic ray impacts are orders of magnitude higher than those of NIS generated phonons, this difference only lasts for a very short time as the high energy phonons relax very quickly~\cite{martinis2021saving}. Eventually, both sources produce phonons with energies sightly above the superconducting gap and are therefore expected to be equivalent with respect to propagation in the chip.

For phonon detection we used microwave kinetic inductance detectors (MKIDs)~\cite{zmuidzinas2012superconducting}. These devices are widely used for the detection of photons or pair-breaking substrate phonons~\cite{day2003broadband,mazin2009microwave}. Their operating mechanism relies on the fact that in superconducting transmission line resonators, in addition to the energy stored in the electromagnetic field, a significant portion of the mode energy is stored in the kinetic energy of the Cooper pairs. This can be modeled by assigning a kinetic inductance $L_k$ to the resonator, so the resonant frequency is 
\begin{equation}
    \omega_0=\frac{1}{\sqrt{C(L_g+L_k)}},
\end{equation}
where $C$ is the total capacitance and $L_g$ is the geometric (electromagnetic) inductance. When a phonon with energy greater than $2\Delta$ is absorbed by the MKID, $\Delta$ being the superconducting gap energy, a Cooper pair is broken which increases $L_k$  which in turn decreases $f_0$. This frequency shift is detected and is indicative of the instantaneous density of QPs in the MKID~\cite{mazin2005microwave,gao2008physics}.

\begin{figure}[htp!]
    \centering
    \includegraphics[width=\linewidth, trim=0 0 0 0, clip]{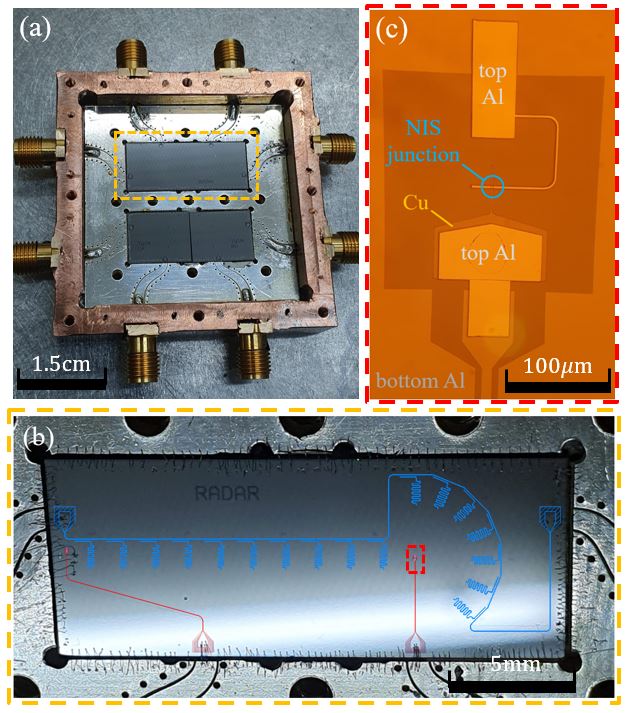}
    \caption{(a) Image of the PCB and copper box on which the chip was mounted, with the lid off. The current experiment was done on the chip in the bottom slot. (b) Image of the chip mounted on the PCB with wirebonds, superimposed with a schematic drawing of the layout. In blue are the MKIDs and their feedline, in red are the NIS junctions and their drive lines. The MKIDs laid out in a circle on the right where not used in this work. (c) A micrograph of the right NIS junction. On top of the silicon substrate it is possible to see the bottom aluminum layer, the copper islands, and the top aluminum layer which creates the junction.}
    \label{fig:chipLayout}
\end{figure}

Our measurements consisted of first resonantly driving the MKIDs so they become loaded with a large number of photons while still keeping them in the linear regime, and then applying a NIS pulse and measuring the IQ plane response. The MKIDs were positioned on a line at increasing distances from the NIS, and were all capacitively coupled to a common feedline in a hanging configuration, as shown in Fig.~\ref{fig:chipLayout}b. For impedance matched hanging resonators the transmission lineshape is given by~\cite{probst2015efficient} 
\begin{equation}\label{eq:S21}
    S_{21}(\omega)=1-\frac{\kappa_e}{\kappa_e+\kappa_i+2 i\left(\omega-\omega_0\right)},
\end{equation}
where $\kappa_e$ and $\kappa_i$ are the external and internal linewidths, respectively.

\begin{figure}[htp!]
    \centering
    \includegraphics[width=\linewidth, trim=0 0 0 10, clip]{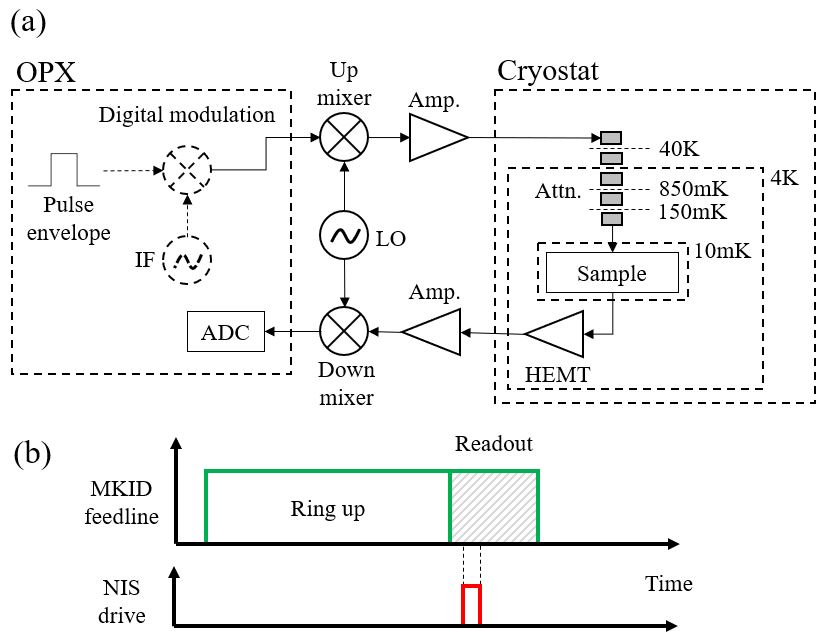}
    \caption{ (a) Signal readout circuit schematic. (b) Pulse sequence used for time domain measurements.}
    \label{fig:readoutScheme}
\end{figure}

To generate the pulse envelopes both for the NIS driving and the MKID readout, we used an OPX+ by Quantum Machines. The raw pulses were digitally modulated by the OPX+ with a 50MHz intermediate frequency (IF) carrier in order to suppress low frequency noise. They were then up-converted with an appropriate local oscillator (LO) to the resonant frequency of the MKIDs and injected into the cryostat. The returning signal was amplified and down-converted with the same LO, and subsequently sampled and recorded by the OPX+ with a 1GHz 12bit analog to digital converter. The signal was digitally demodulated to remove the IF carrier and obtain its in-phase (I) and quadrature (Q) amplitudes. The data was then transferred to a work station for further analyses. The measurement setup is presented in Fig.~\ref{fig:readoutScheme}.

For the MKIDs fabrication we performed electron beam evaporation of an aluminum layer directly on an intrinsic silicon wafer. We then patterned the desired shapes with a Tetramethylammonium hydroxide (TMAH) wet etch process. A thickness of 11nm was measured in an atomic force microscope by Asylum. The NIS junctions were added in a two step process: First, we evaporated 50nm copper islands and patterned them in a liftoff process; Second, in another liftoff process, we opened resist windows for the top aluminum leads and evaporated a 1nm aluminum layer which was completely oxidized to act as the insulating layer of the NIS junction. On top of it, without breaking the vacuum, we evaporated a 100nm aluminum layer for the electrical leads and the NIS junction. The wafer was diced into the correct dimensions and mounted on a designated printed circuit board (PCB) which was soldered to an oxygen-free high thermal conductivity (OFHC) copper box~\cite{huang2021microwave}. The box functioned as a cavity to suppress parasitic microwave modes on the chip. Aluminum wirebonds were used to electrically connect the chip to the PCB. The copper box was sealed with an OFHC copper lid to block external thermal photons, and mounted inside the mixing stage of a commercial Bluefors dilution cryostat. The transmission signal through the feedline was amplified with a high electron mobility transistor (HEMT) amplifier, located on the 4K stage of the cryostat.

\begin{table}
\begin{center} 
\caption{Low power parameters of the measured MKIDs. $\omega_0/ 2\pi$ is the resonant frequency and $\kappa_{e}$ and $\kappa_{i}$ are the external and internal linewidths, respectively. D7 was not observed, probably because of an inadvertent short-circuit.}
\begin{tblr}{colspec = {@{}X[l]X[l]X[l]X[l]@{}}}
\hline
\hline
MKID   & $\omega_0/2\pi \;\;\;\;\;$ [GHz] & $\kappa_{i}/2\pi\;\;\;\;\;\;$ [kHz] & $\kappa_{e}/2\pi\;\;\;\;$ [kHz] \\
\hline
D1     & 3.32522       & 5.1       & 139.2  \\
D2     & 3.22646       & 1.9       & 141.4  \\
D3     & 3.27607       & 4.4       & 70.9   \\
D4     & 3.38292       & 56.5      & 373.4  \\
D5     & 3.64326       & 3.8       & 14.5   \\
D6     & 3.68722       & 2.7       & 13.6   \\
D7     & -             & -         & -      \\ 
D8     & 3.51869       & 13.0      & 108.6  \\
D9     & 3.56871       & 6.0       & 10.3   \\
D10    & 3.46191       & 64.0      & 231.0  \\
\hline
\hline
\label{table:MKIDsParams}
\end{tblr}
\end{center}
\end{table}

\begin{figure}[htp!]
    \centering
    \includegraphics[width=\linewidth, trim=0 20 0 20, clip]{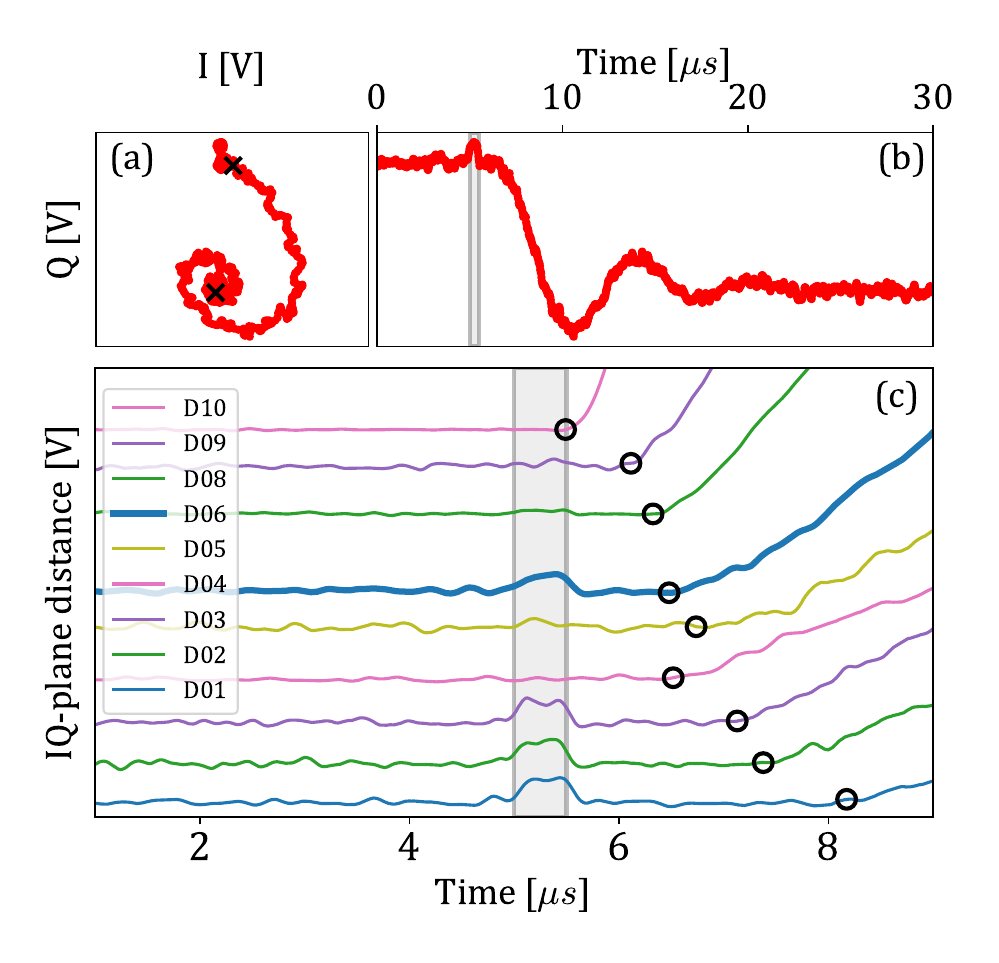}
    \caption{Trajectories of the readout signals after a NIS pulse. (a) The IQ-plane trajectory of the readout signal of D06. The initial and final states are marked by exes. (b) The time evolution of the Q-component of the trajectory in (a), with the NIS pulse time shaded by gray. (c) The time evolution of the IQ-plane distance between the initial and current state of all the readout signals. To improve visibility, traces are shifted vertically in proportion to their distance from the NIS. Burst detection times were found by intersecting a polynomial fit to the post-burst rise with the pre-burst average value. The extracted times are marked by circles.}
    \label{fig:time_trace}
\end{figure}

We characterized the MKIDs with frequency domain measurements using a network analyzer by Keysight. The resonance frequencies and linewidths of the MKIDs that we used in this work are listed in Table~\ref{table:MKIDsParams}. For the time domain measurements we used the pulse sequence in Fig.\ref{fig:readoutScheme}(b). A ring-up pulse of $1\mathrm{ms}$, which is much longer than $1/\kappa$, was first applied to initialize the state of the measured MKID. The readout pulse was juxtaposed immediately afterwards as a direct continuation of the ring-up pulse, with the same amplitude and phase. During the readout pulse, a drive pulse was applied to the NIS, simulating the burst. A typical time trace of the IQ trajectory of the transmission signal from one MKID is presented in Fig.~\ref{fig:time_trace}a,b. From these traces we determined the detection times of the burst at the different MKIDs, which we denote $t_{\mathrm{d}}$. We did this by finding the moment when the IQ plane position of the signal started to deviate from its initial mean value. An automated script was used to accomplish this, together with manual verification to exclude any outliers. To describe the dependence of $t_{\mathrm{d}}$ on the MKID distance and NIS drive power, we developed a simplified model which we present in the following paragraphs. Notice that $t_{\mathrm{d}}$ is only affected by the initial phonons that are emitted form the NIS, so the drive pulse duration has little effect on it.

When a voltage $V_{\mathrm{NIS}}$ is applied to the NIS junction it generates a QP current in the superconductor with maximal energy $eV_{\mathrm{NIS}}$. These QPs slowly diffuse and eventually scatter or recombine to emit phonons. Phonons in the superconductor are efficiently transferred into the substrate, due to the low thickness of the superconductor relative to that of the substrate~\cite{kaplan1979acoustic}, and propagate there. Since the diffusion length of QPs before recombination is a few microns~\cite{kaplan1976quasiparticle} we assume a point-like phonon source. The total number of phonons with energy in the range $[\varepsilon,\varepsilon+d\varepsilon]$ generated at the NIS per unit time is $w_{\mathrm{ph}}^{\mathrm{NIS}}(\varepsilon)d\varepsilon$. Considering an MKID that subtends an angle $\theta$ from the NIS junction, the total number of pair-braking phonons per unit time that reach it is
\begin{equation}\label{eq:phonon_rate}
    W_{\mathrm{ph}}\left(\theta\right)=\left(\frac{\theta}{2\pi}\right) \int_{2\Delta}^{eV_{\mathrm{NIS}}} w_{\mathrm{ph}}^{\mathrm{NIS}}(\varepsilon) d\varepsilon.   
\end{equation}
To get a rough empirical model we use a zeroth order approximation and assume that $w_{\mathrm{ph}}^{\mathrm{NIS}}(\varepsilon)$ is independent $\varepsilon$. Assuming all the electrical power absorbed by the NIS is converted into phonons, we find
\begin{equation}
    \int_0^{eV_{\mathrm{NIS}}} \varepsilon \; w_{\mathrm{ph}}^{\mathrm{NIS}} d\varepsilon = \frac{w_{\mathrm{ph}}^{\mathrm{NIS}} e^2 V_{\mathrm{NIS}}^2}{2} \approx \frac{V_{\mathrm{NIS}}^2}{R_{\mathrm{N}}},
\end{equation}
where $R_{\mathrm{N}}$ is the room-temperature resistance of the NIS junction. We see that $w_{\mathrm{ph}}^{\mathrm{NIS}}=2/\left(e^2 R_{\mathrm{N}}\right)$.  Approximating $\theta\approx L/x$, where $L$ is the active length of the MKID and $x$ is its distance from the NIS, we obtain
\begin{equation}
    W_{\mathrm{ph}}\left(x\right)=\left(\frac{L}{2\pi x}\right) \frac{2 V_{\mathrm{NIS}}}{eR_{\mathrm{N}}} \left( 1 - \frac{2\Delta}{eV_{\mathrm{NIS}}}\right).    
\end{equation}
For $2\Delta \ll eV_{\mathrm{NIS}}$ the last term is close to unity. The time it takes for the burst phonons to reach the MKID is $t_{\mathrm{ph}}=x/v_{\mathrm{ph}}$, where $v_{\mathrm{ph}}$ is the phonon propagation velocity in the substrate. However, the measurement has an additional delay due to the time it takes the MKID to become sufficiently populated by QPs. This delay is  $t_{\mathrm{qp}}=N_{\mathrm{qp}}^0/\eta W_{\mathrm{ph}}$, where $N_{\mathrm{qp}}^0$ is the minimum detectable number of QPs and  $\eta$ is the number conversion efficiency of phonons in the substrate to QPs in the superconductor. The total detection time is therefore
\begin{equation}\label{eq:burst_times}
    t_{\mathrm{d}}=t_{\mathrm{ph}}+t_{\mathrm{qp}}=\frac{x}{v_{\mathrm{ph}}}+\left(\frac{2\pi}{L}\right)\frac{N_{\mathrm{qp}}^0 eR_{\mathrm{N}}}{\eta}\frac{x}{V_{\mathrm{NIS}}}.
\end{equation}
Note that Eq.~\ref{eq:burst_times} describes an apparent burst velocity which is a parallel addition of $v_{\mathrm{ph}}$ and a "parasitic" velocity which depends on the NIS parameters and $V_{\mathrm{NIS}}$. Consequently, we define the velocity that is calculated from $t_{\mathrm{d}}$ as the apparent velocity, and expect it to always be smaller than $v_{\mathrm{ph}}$.\\

To estimate the powers of the drive and readout pulses we used the incoming power into the cryostat and the nominal attenuation of the lines, including reflection from the impedance mismatch at the chip entrance. For the readout pulses, this amounts to about $P_{\mathrm{in}}=1\mathrm{nW}$ in the feedline. The energy stored in the MKIDs is $W_{\mathrm{res}}=2\kappa_i/(\kappa_i+\kappa_e)^2P_{\mathrm{in}}$~\cite{burnett2013high}, which is equivalent to an average photon number of 100 to 20k, depending in the MKID.
For the NIS drive pulses, $P_{\mathrm{NIS}}$, assuming that the transmission line impedance is $Z_0=50\Omega$, the NIS can be well-described as a disconnect. The voltage amplitude on it can therefore be calculated using $V_{\mathrm{NIS}}=2\sqrt{Z_0P_{\mathrm{NIS}}}$.

\begin{figure}[htp!]
    \centering
    \includegraphics[width=\linewidth, trim=0 10 0 0, clip]{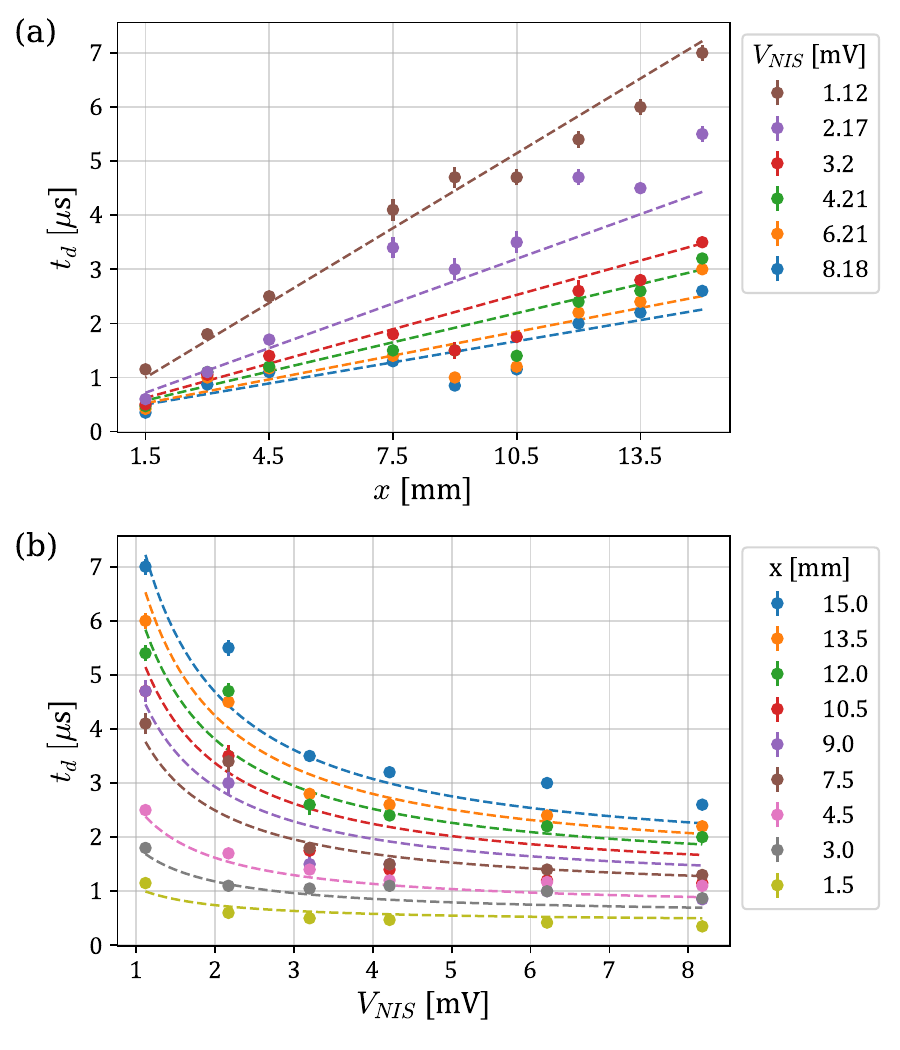}
    \caption{Burst detection times for different MKIDs and NIS drive powers. In (a) each color represents a different drive power, while in (b) each color represents a different MKID. The displayed fits were done over all the data simultaneously using Eq.~\ref{eq:burst_times}. }
    \label{fig:tb_vs_dist_and_Vnis}
\end{figure}

Measurements of the burst detection times at the different MKIDs for various NIS drive powers are shown in Fig.~\ref{fig:tb_vs_dist_and_Vnis}. The detection times were measured relative to the drive pulse start time. A simultaneous fit of all the data to Eq.~\ref{eq:burst_times} is also shown. The extrapolated drive pulse start time was $t_0=0.3\pm 0.12\mathrm{\mu s}$.
An increase in the apparent burst velocity with NIS drive power is clearly seen, as expected. For the best fit, we received $v_{\mathrm{ph}}=12.9\pm 2.6\mathrm{mm/\mu s}$, significantly faster than the expected low temperature longitudinal sound velocity in the [100] direction in silicon~\cite{mcskimin1953measurement}, which is about $\mathrm{9.1 mm/\mu s}$. While the overall trend of the data agrees with the fit and model, we do see some deviations beyond the statistical uncertainties of the measurements. These deviations may appear random but are in fact perfectly repeatable. We do not have a full understanding of their origin, but they possibly stem from local hot-spots on the chip which act as additional phonon sources. Another reason could be phonon reflections from the chip edges which disrupts the linear dependency of the phonon flux with MKID distance. It is worth noting that the deviations do not correlate with the linewidths of the MKIDs.

\begin{figure}[htp!]
    \centering
    \includegraphics[width=\linewidth, trim=0 10 0 12, clip]{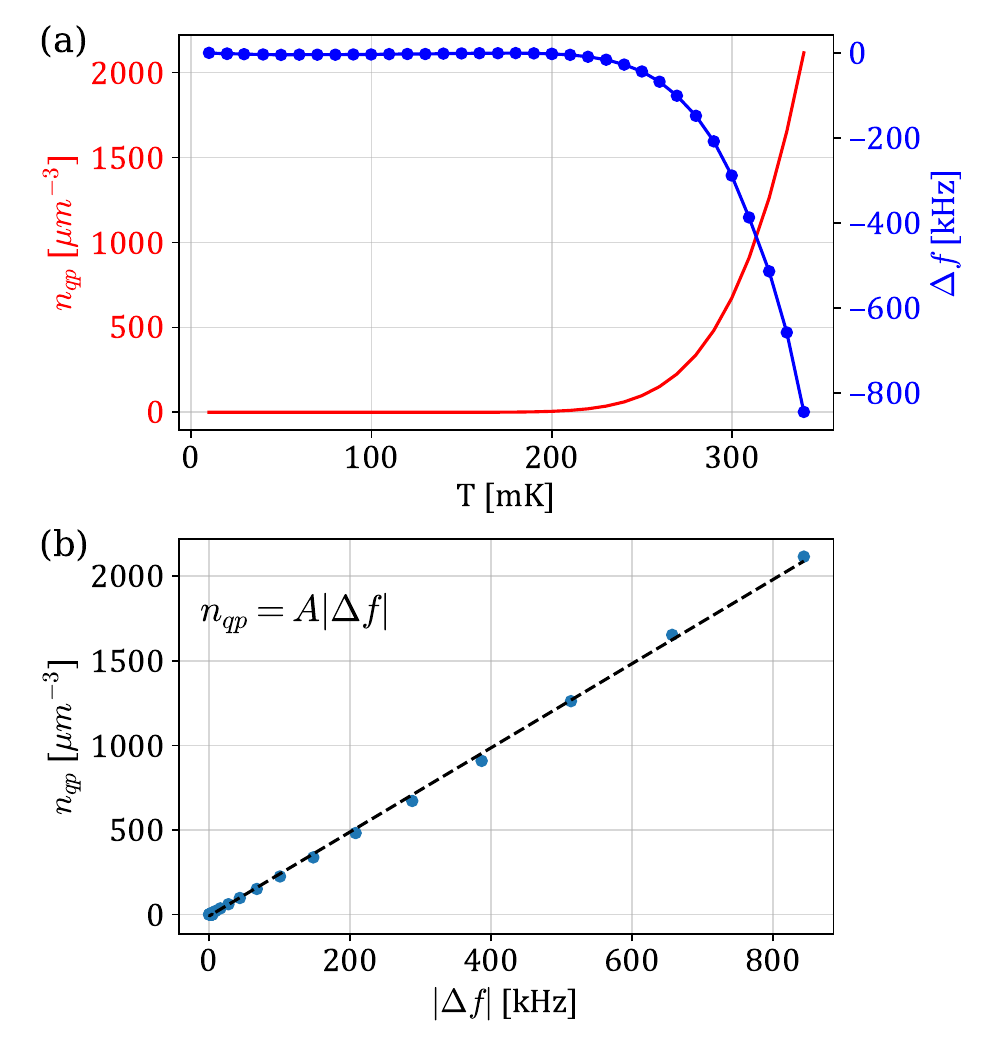}
    \caption{(a) Measured resonance frequency shift with temperature (blue) and theoretical change in QP density according to Eq. \ref{eq:qp_vs_temp} (red). All MKIDs showed similar behavior, so only the results for D01 are shown. (b) Resulting relation between QP density and resonance frequency shift. It is well-described by a linear relation with $A=2.49\mathrm{\mu m^{-3}kHz^{-1}}$, shown as a dashed black line.}
    \label{fig:qp_vs_temp}
\end{figure}

To relate the QP density in the MKIDs to their resonance frequencies we varied the temperature and measured the resulting shift in the resonance frequencies. For each temperature we waited 30 minutes for the system to thermalize. By comparing the results with the well-known formula derived from the Mattis-Bardeen theory~\cite{tinkham2004introduction,mattis1958theory},
\begin{equation}\label{eq:qp_vs_temp} 
    n_{\mathrm{qp}}=2N_0 \sqrt{2\pi k_{\mathrm{B}} T \Delta}\exp^{-\frac{\Delta}{k_{\mathrm{B}} T}},
\end{equation}
we found a linear relation between the QP density, $n_{\mathrm{qp}}$, and the resonance frequency shift, as shown in Fig. \ref{fig:qp_vs_temp}. We used $N_0=1.72\cdot 10^{10}\;\mathrm{\mu m^{-3}eV^{-1}}$~\cite{gao2008equivalence} as the single spin density of states at the Fermi level for aluminum, and $\Delta=238\mathrm{\mu eV}$ as the superconducting gap energy. We obtained $\Delta$ from a measurement of $T_{\mathrm{c}}$ of an aluminum layer from the same evaporation batch by using the BCS relation $\Delta=1.76 k_{\mathrm{B}} T_{\mathrm{c}}$, in accordance with ~\cite{marchegiani2022quasiparticles}. The linear relation between $n_{\mathrm{qp}}$ and the frequency change is well-established in the literature~\cite{fischer2024nonequilibrium}. The expected proportionality factor is roughly $A\approx N_0 \Delta / f_0 \approx 1\mathrm{\mu m^{-3}kHz^{-1}}$, which matches our result of $A=2.49\mathrm{\mu m^{-3}kHz^{-1}}$, as presented in Fig.~\ref{fig:qp_vs_temp}(b). The minimal detectable frequency shift is approximately 1kHz, since this is the scale of the linewidth of our resonators. We can therefore estimate $N_{\mathrm{qp}}^0/\mathcal{V} \approx 2.5\;\mathrm{\mu m}^{-3}$ ($\mathcal{V}$ is the MKID volume) from the linear fit in Fig.~\ref{fig:qp_vs_temp}(b). Note that this estimation is independent of the $t_{\mathrm{d}}$ measurements. The fit of Eq.~\ref{eq:burst_times} gives a value for the pre-factor of the $x/V_{\mathrm{NIS}}$ term, which can be used to estimate $\eta$ since all the other parameters are known. The measured room-temperature value $R_{\mathrm{N}}=10\mathrm{k\Omega}$ and we can approximate $L\approx 3.8\mathrm{mm}$ as the full length of the MKIDs and $20\mathrm{\mu m}$ as their width. We obtained $\eta = 1.3\%$, which can be used to test microscopic models of this coupling.\\

In conclusion, we measured the propagation velocity of high energy phonons in a cryogenicaly cooled silicon chip using MKIDs. The phonon injection source was a NIS junction, which allowed us to achieve temporal resolution of a few nanoseconds and, by repeating the measurements, an arbitrarily good SNR. The results show a clear increase in the apparent burst velocity with increased drive power. We explain this by the combined effect of the existence of a detection threshold for the MKIDs, and the decay of the phonon flux with the distance from the source. We developed a simplified model which captures this effect, and although our measurements showed some deviations from it, the agreement is encouraging. A possible improvement would be a more accurate estimation of the NIS drive power, which in the current experimental system had to be approximated as the nominal value. The effect could explain the low apparent propagation velocity measured in~\cite{Swenson2010highSpeed}, and might be used to deduce the burst energy from it. Knowing the error threshold for an element on a chip, our model can give the propagation velocity of the errors due to a burst, which is important for knowing the error rate and range in the chip. \\

This research was supported by the Israeli Science
Foundation (ISF), Israel Ministry of Science and Technology (MOST), and Technion’s Helen Diller Quantum Center. The data that support the findings of this study are available from the corresponding author upon reasonable request.

\bibliographystyle{apsrev4-2}
\bibliography{bib}

\setcounter{equation}{0}
\setcounter{figure}{0}
\setcounter{table}{0}
\setcounter{page}{1}
\makeatletter
\renewcommand{\theequation}{S\arabic{equation}}
\renewcommand{\thefigure}{S\arabic{figure}}
\renewcommand{\bibnumfmt}[1]{[S#1]}
\renewcommand{\citenumfont}[1]{S#1}

\onecolumngrid

\end{document}